\documentclass[pdftex,a4paper]{scrartcl}
\pdfoutput=1
\usepackage[ngerman,english]{babel}
\useshorthands{"}
\addto\extrasenglish{\languageshorthands{ngerman}}
\usepackage[utf8]{inputenc}
\usepackage[T1]{fontenc}
\usepackage{amsmath}
\usepackage{graphicx}
\usepackage{amssymb}
\usepackage{hyperref}
\usepackage{mathrsfs}
\usepackage{color}

\renewcommand{\phi}{\varphi}
\renewcommand{\theta}{\vartheta}

\title{{Goldstone origin of black hole hair from supertranslations and criticality}}
\author{{\bf Artem Averin$^{\textrm{a,b}}$, Gia Dvali$^{\textrm{a,b,c}}$, Cesar Gomez$^{\textrm{d}}$, Dieter L\"ust$^{\textrm{a,b}}$}}

\begin{document}

\maketitle

\centerline{\it $^{\textrm{a}}$ Arnold--Sommerfeld--Center for Theoretical Physics,}
\centerline{\it Ludwig--Maximilians--Universit\"at, 80333 M\"unchen, Germany}
\medskip
\centerline{\it $^{\textrm{b}}$ Max--Planck--Institut f\"ur Physik,
Werner--Heisenberg--Institut,}
\centerline{\it 80805 M\"unchen, Germany}
\medskip
\centerline{\it $^{\textrm{c}}$ Center for Cosmology and Particle Physics,
Department of Physics, New York University}
\centerline{\it 4 Washington Place, New York, NY 10003, USA}
\medskip
\centerline{\it $^{\textrm{d}}$ Instituto de F\'{\i}sica Te\'orica UAM-CSIC, C-XVI,
Universidad Aut\'onoma de Madrid,}
\centerline{\it Cantoblanco, 28049 Madrid, Spain}

\vskip1cm
\abstract{{
Degrees of freedom that carry black hole entropy and hair can
 be described in the language of Goldstone  phenomenon. 
  They represent the pseudo-Goldstone bosons of certain supertranslations, called 
  ${\cal A}$-transformations,  that 
  are spontaneously broken by the black hole metric. This breaking gives rise to a  
  tower of Goldstone bosons created by the spontaneously-broken
  generators that can be labeled by spherical harmonics. Classically, the number of charges is infinite, they have vanishing 
 VEVs and the corresponding Goldstone modes are gapless.  The resulting hair and entropy are infinite, but unresolvable. 
   In quantum theory the two things happen. The number of legitimate Goldstone modes 
   restricted by requirement of weak-coupling, becomes finite and scales  as black hole area in Planck units. The Goldstones generate a tiny gap, controlled by their gravitational coupling. The gap turns out to be equal to the inverse of black hole half-life, $t_{BH}$. Correspondingly,  in quantum theory the charges are neither  conserved nor vanish, but non-conservation time is set by $t_{BH}$.  
   This picture nicely matches with the idea of a black hole as of critical system composed of many soft gravitons. The $\cal A$-Goldstones of geometric picture represent the near-gapless Bogoliubov-Goldstone modes of critical
 soft-graviton system. }}

\begin{flushright}
{\small  MPP--2016--139},
{\small LMU--ASC 27/16}
\end{flushright}

\newpage

\setcounter{tocdepth}{2}
\break

 The key target of this note lies in understanding the black hole entropy in terms of 
 Goldstone-type phenomenon in which the black hole hair and entropy are encoded 
 in Goldstone modes of symmetries that are spontaneously-broken by the black hole geometry.  
     This spontaneous breaking gives rise to a set of gapless Goldstone modes
     which have the right properties for being the promising candidates for carrying the black hole hair and entropy. \\ 
 
 The Goldstone effect is an universal phenomenon of nature that gives rise to 
 gapless excitations, the Nambu-Goldstone modes. 
 The necessary condition for their appearance is that the transformation that acts 
 on the vacuum non-trivially is physical, i.e., is not a gauge redundancy. \\  
 
  {\it Thus, for explaining phenomena that demand existence of 
 gapless modes, it is natural to look for their origin in  an underlying Goldstone mechanism.}  \\
 
   This idea was applied to black holes in \cite{critical}.  Indeed, a black hole of  Schwarz\-schild radius  $r_S$ carries Bekenstein-Hawking entropy, 
  \begin{equation} 
  \label{entropy} 
    S \,  = \,  {r_S^2 \over L_P^2} \, ,
  \end{equation} 
  where $L_P= \sqrt{\hbar G_N}$ is the Planck length and $G_N$ is Newton's constant (Here and below we ignore numerical factors whenever they are not necessary).   
Obviously,  the explanation of this entropy demands the existence of large number of 
nearly-gapless degrees of freedom.  
 
  In \cite{critical} these were identified with Goldstone-type modes of black hole 
  ground-state. In this
 picture the black hole is described as $N$-graviton system \cite{Nportrait} at the quantum critical point at which the graviton number and quantum gravitational coupling $\alpha$ satisfy the condition $\alpha N = 1$.  
 The Goldstone modes that carry black hole entropy are the collective Bogoliubov modes of critical multi-graviton system \cite{critical}.  The mapping between the Goldstone and 
 Bogoliubov modes is very transparent in many-body description \cite{gold}. 
 It was shown that Bogoliubov modes of a critical system are the  
 Goldstone modes of spontaneously broken symmetry of the system. 
 
  In quantum theory these modes are not exactly gapless and there exists a finite gap of order $\epsilon = {\hbar \over r_SN}$.  In this description $N$ emerges to be equal to black hole entropy, $S$, because,  due to criticality,  it is equal to the inverse coupling,  $\alpha^{-1}  = 
  {r_S^2 \over L_P^2}$, of gravitons of wavelength $r_S$.   Thus, the information 
  about the black hole entropy is encoded into the quantum gravitational coupling of gravitons that form it.  Because of quantum criticality, the same quantity also describes the number of constituent gravitons needed to form a black hole \cite{critical,Nportrait,other}.   Hence, in what follows, we shall not make any distinction between the quantities $N$ and $S$.
 
  Existence of the finite ${1 \over N}$-gap in Goldstone modes means 
  that the sponta\-neously-broken transformation is not an exact symmetry of the 
  quantum theory. 
  Again, this fact is very transparent in many-body language and is because of 
  ${1\over N}$-corrections - due to quantum fluctuations - that move system away from the critical point.

   The question that was addressed in \cite{Dvali:2015rea,Averin:2016ybl} is:  what is the 
 classical geometric limit of spontaneously broken symmetries that account 
 for black hole entropy and hair?  Notice, since in the classical limit we have $N \rightarrow \infty$, it is expected that in this limit Goldstone modes must become exactly gapless. \\

  In this short note we summarize the basic points in \cite{Dvali:2015rea,Averin:2016ybl} as they have been presented in several conferences by the authors during last few months.
Related ideas appeared in \cite{Hawking:2015qqa,HPS} (see also \cite{many}). 

    The central point of \cite{Averin:2016ybl} can be summarized as follows:
    for black hole space time the asymptotic BMS-like symmetries are enhanced 
  - as compared to Minkowskian space-times -  by an infinite set of {\it spontaneously broken} transformations, denoted by ${\cal A}$.  The Goldstone modes of these transformations are responsible for  black hole entropy and hair and exist do to criticality of black hole quantum state.  In quantum theory their number is finite and they are no longer exactly gapless.   
  
  Following \cite{Averin:2016ybl}, we  shall identify the proper geometric 
 classical limit ($\hbar \rightarrow 0$)  of symmetries that are spontaneously broken by black hole geometry.  Then - using simple physical argument - 
 we shall provide evidence that: \\
 
 {\it 1) Classically, there are infinite number of gapless Goldstones that can be labeled by spherical harmonics, $lm$.  The corresponding charges,  $Q_{lm}$,  vanish. \\
 
  2)  For finite $\hbar$ (i.e., finite $N$), the number of legitimate Goldstone modes  scales as $N$,  in accordance with Bekenstein-Hawking  entropy. \\

 3) For finite $\hbar$ the charges $Q_{lm}$ acquire expectation values and are no longer conserved. Both of these effects are suppressed  by ${1 \over N}$.
 Corresponding non-conservation time is therefore given by the black hole half life-time, $t_{BH} = N r_S$. \\  
 }

 In  \cite{Averin:2016ybl}, first we have derived the specific form of the supertranslations that leave certain boundary and coordinate transitions on the horizon invariant.   
   In the well-know Schwarzschild metric in Eddington-Finkelstein 
   coordinates,\footnote{Here $\gamma_{z\bar z}={2\over(1+z\bar z)^2}$ is the round metric and the complex coordinates are defined as $z=\cot\bigl({1\over 2}\theta\bigr)e^{i\phi}$
and the metric of $S^2$ in angular coordinates reads $ds^2=2r^2\gamma_{z\bar z}dzd\bar z=d\theta^2+\sin^2\theta d\phi^2$.}  
\begin{equation}\label{schwarzschildmetric}
ds^2 = -(1-\frac{r_S}{r})dv^2 + 2dvdr + r^2 \gamma_{z\bar z}dzd\bar z\,, 
\end{equation}  
 they act on the coordinates $(v,r,z,\bar z)$ in the following way:
\begin{eqnarray}\label{horizontrans}
 BMS_H:\quad v&\rightarrow& v-f(z,\bar z)\, ,\nonumber\\
z&\rightarrow &z+{r_S-r\over rr_S}\gamma^{z\bar z}\partial_{\bar z}f(z,\bar z)\, ,\nonumber\\
\bar z&\rightarrow &\bar z+{r_S-r\over rr_S}\gamma^{z\bar z}\partial_{ z}f(z,\bar z)\, ,\nonumber\\
r&\rightarrow &r\, .
\end{eqnarray}

Second, we have identified a candidate transformation that can be a source 
for Goldstone modes that carry black hole hair and  which was referred to as $\cal A$-transformation \cite{Averin:2016ybl}.
 The ${\cal A}$-transformations change the metric 
(\ref{schwarzschildmetric}) in the following way:
\begin{align}
{\cal A}:\quad\delta_{\chi_f}g_{\mu \nu}=
\begin{pmatrix}
0 & 0 & 0 & 0\\
0 & 0 & 0 & 0\\
0& 0 & -2r^2 \frac{1}{r_S} \frac{\partial^2 f}{\partial \theta^2} & -2r^2 \frac{1}{r_S} (\frac{\partial^2 f}{\partial \theta \partial \phi}-\cot \theta \frac{\partial f}{\partial \phi})\\
0 & 0 & * & -2r^2 \frac{1}{r_S} (\frac{\partial^2 f}{\partial  \phi^2} + \sin \theta \cos \theta \frac{\partial f}{\partial \theta}).
\end{pmatrix}
\label{microstates}
\end{align}  
These transformations are of BMS-type (see \cite{Ashtekar} for a review on asymptotic symmetries).   
In the limit of large black holes we can think of them by formally substracting from  
  $BMS_H$ the action of the standard BMS on the null infinity.
One can in addition show that 
 the metric $g_{\mu\nu}^f=g_{\mu\nu}+\delta_{\chi_f}g_{\mu \nu}$  is Ricci-flat  in the angular directions:
 \begin{equation}
 R_{\theta\phi}(g_{\mu\nu}^f)=R_{\theta\theta}(g_{\mu\nu}^f)=R_{\phi\phi}(g_{\mu\nu}^f)=0\, .
 \end{equation}

  Expanding the Schwarzschild metric as 
\begin{eqnarray}
ds^2 &=& -(1-\frac{r_S}{r})dv^2 + 2dvdr + r^2 \gamma_{z\bar z}dzd\bar z\nonumber\\
&+& {r_S}C^{{\cal A}}_{zz} dz^2+{r_S}\bar C^{{\cal A}}_{\bar z\bar z}d\bar z^2+\dots \, .
\,,\label{cametric}
\end{eqnarray} 
one can equivalently express the  ${\cal A}$-transformations eq.(\ref{microstates}) as transformations on 
the
gauge connections $C_{zz}^{{\cal A}}$ as:  
\begin{equation}\label{cHtrans}
{\cal A}:\quad C^{{\cal A}}_{zz}\,\rightarrow \, C^{{\cal A}}_{zz}-2D^2_zf\, .
\end{equation}
The functions $2D^2_zf$ now correspond to the infinite family of metric variations in eq.(\ref{microstates}).

  The crucial point for the further discussion is that these transformations 
   are defined by arbitrary functions of angular variables $f(\theta,\phi)$. 
    This means that classically there exist infinite number of generators
  that are {\it spontaneously broken}  by the black hole geometry.  These generators 
  $Q_{lm}$ can be labelled by the spherical harmonics, $l\, m$.

   Correspondingly to these charges, there exist 
  {\it infinite } number of independent Goldstone modes that can be created by the action on a black hole vacuum by these generators. 
 In this labeling the given Goldstone mode $b_{lm}$ 
 is created by a charge $Q_{lm}$ that represents a generator of the  $\cal A$-transformation 
 (\ref{microstates})  for a given spherical harmonic  $f_{lm}$ of the function
 $f(\theta,\phi) \, = \, \sum_{l,m} \, f_{ml}\, Y_{lm}(\theta,\phi)$.

 In quantum language, describing an unperturbed classical metric (\ref{schwarzschildmetric}) by a black hole vacuum state $|BH\rangle$, we can write
   \begin{equation}
  Q_{lm} |BH\rangle = |BH\rangle_{lm} \, .
 \end{equation}
Since the classical  $\cal A$-transformations preserve the ADM mass, it is clear that 
the corresponding Goldstone modes are all gapless, i.e.,  they have zero frequencies 
$\omega_{lm} = 0$. Notice, this  {\it does not} mean that their wavelengths are infinite, 
since these modes do not satisfy the dispersion relations of free propagating waves on a flat space-time.
In fact, the near horizon  expansion of the metric in eq.(\ref{cametric})  indicates that the radial wave-lengths of the relevant  ${\cal A}$-modes are of the  order of the horizon, i.e. $\lambda_{\cal A}\sim r_S$.
 Due to zero gap of Goldstones  modes, the expectation values of charges over classical 
black hole vacua vanish.\footnote{The fact that charges vanish in the classical vacuum can be directly seen from their classical expression. 
 $ Q =  \int \bigl(g^{\mu\nu} \partial_v\delta_{\chi_f}g_{\mu \nu}\bigr)=
  \int ~\partial_v(D^2_zf)+h.c.$, where the integration contour  also contains the horizon of the black hole.

   Alternatively, the classical charges can be obtained as $\omega_{lm} = 0$ limit of the quantum expression 
$Q_{lm} \, = \, -i  \, \sqrt{\omega_{lm}\hbar}  \left(
\hat b_{lm} e^{-iv\omega_{lm}} \, + \, \hat b^{\dagger}_{lm}e^{iv\omega_{lm}} \right)$, 
where  $ \hat b^{\dagger}_{ml}, \hat b_{ml}$ are creation and annihilation operators of different spherical modes.   The classical vacua can be constructed as coherent states obtained by the action of displacement operator
at the unperturbed black hole vacuum, 
$e^{i\epsilon_{lm} Q_{lm}} |BH\rangle \, = |coh\rangle_{lm}\,$.
 It is clear that the classical values of $Q_{lm}$ vanish.}

  Thus, from above it is clear that in classical theory ($N = \infty,  \hbar = 0$) the  black hole carries infinite hair and entropy, but 
 the resolution time is also infinite.  What happens in the quantum theory?

  We wish to show that in quantum theory the number of ``legitimate" Gold\-stone modes  is finite and 
 scales as $N$.  The criterion of legitimacy is very simple: in quantum theory 
 we shall only keep the modes which are {\it weakly-coupled}.
 We thus have to estimate the number of such modes. 

    The number of independent charges and of corresponding Goldstone modes 
  below a given level $l$ scales as  $N_{gold} \sim l^2$.  In classical theory 
  their number is infinite, since $l$ can take an arbitrarily-large value without 
  any restriction.  However, in quantum theory only the modes up to certain $l_{max}$ can be included. 
   
  We need a quantum criterion for determining $l_{max}$. As said above, the criterion is that we can only include the set of Goldstone modes that are weakly-coupled. 
   The quantum gravitational coupling of generic modes of angular number $\sim l$
   and a radial wavelength $\lambda_r$ goes as  $\alpha =  {l^2 L_P^2\over \lambda_r^2}$.   
    We thus determine 
   $l_{max}$ by imposing the condition of weak coupling,  $\alpha_{max} = 1$. 
   
   Since we are interested in near-horizon physics, from the form of $\cal A$-transfor\-ma\-tions it is clear that we should take the radial wavelength of the Goldstone modes to be $\lambda_r \sim r_S$.   
   This determines the maximal harmonic number as $l_{max} = {r_s \over L_P}$, and thus,  the resulting number of legitimate Goldstone modes scales as, 
     \begin{equation}
         N_{gold} \sim l_{max}^2  \sim N  \, .
         \label{entropy} 
   \end{equation}  
 Thus, from a simple physical argument - counting only the number of weakly-coupled modes - we obtained the number of Goldstone species that matches the 
 Bekenstein-Hawking entropy.   
  
   As pointed out in \cite{Averin:2016ybl} in quantum theory these Goldstones are expected to develop a finite energy gap.  For weakly-coupled Goldstones this gap was estimated to be 
   \begin{equation} 
  \epsilon \sim {1 \over N} {\hbar \over r_s} \, .
  \label{gap}  
   \end{equation}
   Notice that the gap is inverse of black hole half life-time $t_{BH} \sim \hbar \epsilon^{-1}$.   
 The estimate relied on microscopic information from quantum criticality, which suggests that for a given mode with gravitational self-coupling $\alpha$, the 
 gap must scales as $\epsilon \sim {\alpha^2 \over N}$. 
  However, the inevitability of the energy gap can already be guessed from the classical 
  theory. Namely, from the fact that the  ${\cal A}$-transformation 
 (\ref{microstates}) - although it preserves the ADM mass of Schwarzschild geometry
 -  does not necessarily preserve the masses  of other classical configurations 
 that can be obtained by arbitrary deformations of Schwarzschild geometry.  
  At the classical level, such deformed geometries cannot generate the 
 gap for Goldstone waves obtained by ${\cal A}$-transforming the Schwarzschild
  background.  However, for $\hbar \neq 0$, the virtual processes to which such classical deformations contribute will in general lift the degeneracy  of ${\cal A}$-Goldstone vacua and create an energy gap. 
  Putting it simply,  the ${\cal A}$-transformations - despite preserving the black hole mass - are not symmetries of the full action.  Due to this,  the degeneracy of 
  vacua must  be lifted by quantum effects and the Goldstone modes must acquire the finite energy gaps.  
  
   Since by  usual Goldstone relation the energy gap in the Goldstone mode controls the non-conservation of the corresponding charge, we can estimate that
   the charges $Q_{lm}$ are not  conserved on the time scale of $t_{BH} = \hbar \epsilon^{-1}$.  
   Correspondingly, they shall acquire non-zero VEVs,
    \begin{equation}
 \langle BH | Q_{lm} |BH\rangle \sim {1 \over N} \, , 
 \label{vev}
 \end{equation}
that shall evolve over the time-scale $t_{BH}$.

The above expression must be taken with a great care.  
In fact, it is pointless to look for more precise expression  
without being able to take into the account the effect of black hole evaporation and other quantum phenomena. 
  This is because, the VEVs are tiny and evolve on the time-scale 
  comparable with the black hole life-time. 
  During this time-scale half of the black hole has already evaporated and the quantum mechanics has order-one effect.   
  
   In other words, it is certainly true that the ${\cal A}$-charges acquire expectation values, but the time-scale to measure them is comparable with black hole life-time
   and one has to take into the account the back reaction from evaporation. 
   All these facts consistently indicate that the time-scale for resolving the black hole hair exceeds $t_{BH}$.

  Finally notice, that for Minkowski space the charges continue to vanish -
  as well as to be conserved - also in the quantum theory. The  simple reason for this 
  is that the wavelengths of  Goldstone bosons for Minkowski space are 
  infinite  $\lambda = \infty$. These Goldstone bosons correspond 
  to standard BSM transformations.
     Correspondingly, we have  $l_{max} = \infty$ and there are infinite number of conserved charges with gapless Goldstones.  
   Quantum-mechanically Minkowski vacua represent the coherent states of infinite wavelength gravitons 
   with infinite occupation numbers \cite{Dvali:2015rea}.      
     Putting it differently, even in quantum theory Minkowski carries an infinite hair and infinite entropy, but its resolution requires an infinite time.

 In summary, the charges associated with asymptotic symmetries as well as those for the case of horizons are always quantum mechanically determined by the relevant soft Goldstone modes. In the case of null infinity these can be thought of as  standard radiative soft modes, but of infinite wavelength. In the case of black holes we have identified them with Goldstone modes of 
 ${\cal A}$-transformations, of finite wave-length. In both cases classically there are 
 infinite number of charges, labeled by $lm$,  with vanishing VEVs and  with associated gapless Goldstones. For Minkowski,  this continues to be true 
 also in quantum theory, due to the infinite wavelength of Goldstones.  
 But, for a finite size black hole the story is different, and the number of 
 charges as well as the gap become finite and is controlled by $N$. 
   The resulting time-evolution scale for charges is $t_{BH}$, which shows 
  why the resolution time for black hole hair cannot be shorter than the black hole half-life time.

\end{document}